\documentclass[12pt]{article}
\usepackage{graphicx,psfrag}

\newcommand{\preprint}[1]{\begin{flushright}#1\end{flushright}}

\newcommand{\bea}{\begin{eqnarray}}
\newcommand{\eea}{\end{eqnarray}}
\newcommand{\beq}{\begin{equation}}
\newcommand{\eeq}{\end{equation}}
\newcommand{\simgt}{\hbox{ \raise3pt\hbox to 0pt{$>$}\raise-3pt\hbox{$\sim$} }}
\newcommand{\simlt}{\hbox{ \raise3pt\hbox to 0pt{$<$}\raise-3pt\hbox{$\sim$} }}

\def\a5{${\cal O}(\alpha_S^5 m)$}
\def\la5{${\cal O}(\alpha_S^5 m \log \alpha_S )$}

\begin{document}

\preprint{TU-675\\Nov. 2002}

\vspace*{2cm}
\begin{center}
{\bf \Large  Top mass determination and \boldmath{${\cal O} (\alpha_S^5 m)$} 
correction to\vspace{3mm}\\ 
toponium $1S$ energy level
}
\\[20mm]
{\large Y. Kiyo and Y. Sumino}
                \\[10mm]
{\it Department of Physics, Tohoku University\\
                          Sendai, 980-8578 Japan }
\end{center}
\vspace{3cm}
\begin{abstract}
Recently the full ${\cal O}(\alpha_S^5 m, \alpha_S^5 m \log \alpha_S)$ 
correction to the heavy quarkonium $1S$ energy level has been
computed (except the $a_3$-term in the QCD potential).
We point out that the full correction (including the $\log \alpha_S$-term)
is approximated well by the large-$\beta_0$ approximation.
Based on the assumption that this feature holds up to higher orders, 
we discuss why the top quark pole mass cannot be
determined to better than ${\cal O}(\Lambda_{\rm QCD})$ accuracy
at a future $e^+e^-$ collider, while the $\overline{\rm MS}$ mass
can be determined to about 40~MeV accuracy 
(provided the 4-loop $\overline{\rm MS}$-pole mass
relation will be computed in due time).

\end{abstract}

\newpage

Recently a large part of the \a5 corrections \cite{nnnlo-H, Penin-Steinhauser} 
to the energy spectrum
of the heavy quarkonium $1S$ state has been calculated.
Combining this with the previously known \la5 corrections 
\cite{Brambilla-Pineda-Soto-Vairo, Kniehl-Penin}
the only remaining piece to be computed in order to complete the \a5
corrections is the non-logarithmic term ($a_3$) of the static QCD 
potential at 3-loop.
Using a Pad\'e estimate \cite{Chishtie-Elias} of $a_3$, 
Ref.~\cite{Penin-Steinhauser} examined the scale dependences
and the convergence properties of the bottomonium $1S$ and
the (would-be) toponium $1S$ energy levels.
The dependences of the energy levels on the value of $a_3$ are
found to be rather weak.
As for the toponium case,
Ref.~\cite{Penin-Steinhauser} concluded that the top quark pole mass 
can be extracted from the $1S$ energy level with a theoretical 
error of about 80~MeV.
This estimate of the theoretical error on the top quark pole mass
appears to be considerably smaller as compared to a previous
common consensus that the pole mass has a theoretical uncertainty of
order $\Lambda_{\rm QCD}$$\sim 200$--300~MeV \cite{topcollab}.

In this paper we discuss two issues.
First we point out that the presently known \a5 correction to the
$1S$ energy level is approximated fairly well by its
large-$\beta_0$ approximation (naive nonabelianization) \cite{bb}.
We consider this fact to be quite non-trivial because of
the following reason. 
We know that from \a5 the ultrasoft scale starts to contribute
to the energy level. 
Since it is a completely new type of contribution 
(as compared to the lower-order corrections), and since it is 
generally believed to give very large corrections
\cite{Brambilla-Pineda-Soto-Vairo, Kniehl-Penin}, we have expected that the 
large-$\beta_0$ approximation may well fail to be a good approximation at 
\a5 in the energy level.

One may wonder that our point, that the large-$\beta_0$ approximation is
good, is in contradiction to the conclusion of \cite{Penin-Steinhauser}: 
``We have found that the N$^3$LO corrections are dominated neither by 
logarithmically enhanced $\alpha_S^3 \ln(\alpha_S)$ nor by the renormalon 
induced $\beta_0^3 \alpha_S^3$ terms and thus the full calculation of 
the correction is crucial for quantitative analysis."
In fact, there is no contradiction, because the definition of the
``$\beta_0^3 \alpha_S^3$ terms" in \cite{Penin-Steinhauser}
differs from that of the usual large-$\beta_0$ approximation.\footnote{
For instance, the term proportional to $a_1 \beta_0^2$ is not included in the 
$\beta_0^3$ term of \cite{Penin-Steinhauser}, whereas 
a part of $a_1 \beta_0^2$ is included in the large-$\beta_0$ approximation.
}
Nevertheless, 
we have to say that the above statement of \cite{Penin-Steinhauser} is
quite misleading, since it does not address the difference between
its $\beta_0^3 \alpha_S^3$ terms and the large-$\beta_0$ approximation, 
and since it is the large-$\beta_0$ approximation that is the empirically 
successful approximation and, therefore, the renormalon dominance picture has 
often been discussed in this context in the literature.

Secondly, we discuss an error estimate of the top quark pole mass based on the 
assumption that the large-$\beta_0$ approximation continues to be a good approximation 
up to higher orders. At the same time we discuss the accuracy with which the
top quark $\overline{\rm MS}$ mass can be extracted from the toponium $1S$ energy
level.

The sum of the full \a5 and \la5 corrections to the energy level of the
heavy quarkonium $1S$ state is given in 
Eqs.~(6), (12) and (13) of \cite{Penin-Steinhauser}.
The part unrelated to the lower-order
corrections via the renormalization-group equation 
(for the running of the coupling) can be extracted by
setting $L_\mu=\log[ \mu/(C_F \alpha_S(\mu) m_{\rm pole}) ] =0$.
It reads numerically
\bea
&& \hspace{-1cm}
\delta E_1^{(3)}\Bigr|_{L_\mu=0} 
= 
- \frac{(C_F \alpha_S(\mu))^2}{4} \,  m_{\rm pole} \times 
\biggl( \frac{\alpha_S(\mu)}{\pi} \biggr)^3 \, c_3  \, ,
\\ 
&& \hspace{-1cm}
c_3 
\simeq
7078.8 + 0.03125\, a_3 - 1215.5\, n_f + 69.451\,{{n_f}^2} - 1.2147\,{{n_f}^3} 
+  474.29\, \log \left( \alpha_S (\mu )\right) \, ,
\label{full}
\eea
where $C_F=4/3$ is a color factor.

In general, the large-$\beta_0$ approximation of a quantity, at a given order of perturbative
expansion in $\alpha_S$, is defined as follows:
We first compute the leading order contribution in an expansion in
$1/n_f$, where $n_f$ is the number of light quark flavors, which
comes from so-called bubble chain diagrams.
Then we transform this large $n_f$ result by a simplistic replacement
$n_f \to n_f - 33/2 = -(3/2)\beta_0$.
In many phenomenological applications the large-$\beta_0$ approximation 
turns out to be a good approximation of the full result for quantities 
which contain the leading renormalon, see e.g. 
\cite{Beneke,3loop-mass-relation,large-b0-1,large-b0-2}.
The corresponding correction to Eq.~(\ref{full})
in the large-$\beta_0$ approximation is given by \cite{Kiyo-Sumino, Hoang}
\bea
c_3(\mbox{large-}\beta_0) &=& \beta_0^3 \, \Biggl(
{\frac{517}{864}} + {\frac{19\,{{\pi }^2}}{144}} + 
  {\frac{11\,{\zeta_3}}{6}} + 
  {\frac{{{\pi }^4}}{1440}} - 
  {\frac{{{\pi }^2}\,{\zeta_3}}{8}} + 
  {\frac{3\,{\zeta_5}}{2}}
\Biggl)
\nonumber \\
&\simeq& 
5649.36 - 1027.16\,{n_f} + 62.2519\,{{n_f}^2} - 1.25761\,{{n_f}^3} \, .
\eea

In Table~\ref{table:c3} we compare $c_3$ and 
$c_3(\mbox{large-$\beta_0$})$ for values of $n_f$ and $\alpha_S$
corresponding to the $\Upsilon (1S)$ and toponium $1S$ states.
For $a_3$, we used the Pad\'e estimate \cite{Chishtie-Elias} as well as  
the estimate based on the renormalon dominance picture \cite{pinedaJHEP};
$c_3$ differs by less than $3 \%$ when we use these estimates,
for $n_f=4,\, 5$.\footnote{
The corresponding estimates of the three loop coefficient 
$a_3$ are given by
$a_3(\mbox{Pad\'e})/4^3=98, \, 60$, and $a_3(\mbox{Pineda})/4^3=72,\,37$ 
for $n_f=4,\, 5$, respectively  \cite{Chishtie-Elias,pinedaJHEP}.
}
We also varied $a_3$ by $\pm 100$\% in Eq.~(\ref{full}) and find that $c_3$ 
changes by less than $\pm 10$\% for $n_f=4,\, 5$. 
As we can see from the table, the large-$\beta_0$ approximation turns out to 
lie between 85\% and 120\% of the full result in the relevant cases.
We observe that the agreement becomes substantially worse if we
remove the $\log \alpha_S$ term from the full result.
\begin{table}
\begin{center}
\begin{tabular}{l|ccc}
\hline
\makebox[38mm]{$\alpha_S$}&\makebox[38mm]{0.1}&\makebox[38mm]{0.2}&\makebox[38mm]{0.3} \\
\hline
\hline
$c_3~|_{n_f=4}$   & 2354  & 2683  & 2875         \\
$c_3~|_{n_f=5}$   & 1613  & 1942  & 2134           \\
\hline
$c_3~|_{n_f=4,\, \log(\alpha_S)\rightarrow 0}$ 
                  & 3446  & 3446  & 3446          \\
$c_3~|_{n_f=5,\, \log(\alpha_S)\rightarrow 0}$ 
                 & 2705  & 2705  & 2705          \\
\hline
$c_3(\mbox{large-}\beta_0)~|_{n_f=4}$  &   2456 (104\%)  &   2456 (92\%)  &   2456 (85\%)  \\
$c_3(\mbox{large-}\beta_0)~|_{n_f=5}$  &   1913 (119\%)  &   1913 (98\%)  &   1913 (90\%) \\
\hline
\end{tabular}
\caption{\footnotesize 
Numerical values of $c_3$ and its large-$\beta_0$ results are shown 
for $\alpha_S=0.1,\, 0.2,\, 0.3$ and $n_f=4,\,5$. 
For each $c_3(\mbox{large-}\beta_0)$, the ratio to the full result ($c_3$) 
is shown in the parenthesis.
The Pad\' e estimate \cite{Chishtie-Elias}
of $a_3$ is used in Eq. (\ref{full}) to obtain $c_3$.
} 
\label{table:c3}
\end{center}
\end{table}

\begin{figure}
\begin{center}
\begin{tabular}{cc}
\includegraphics[width=6cm,angle=-90]{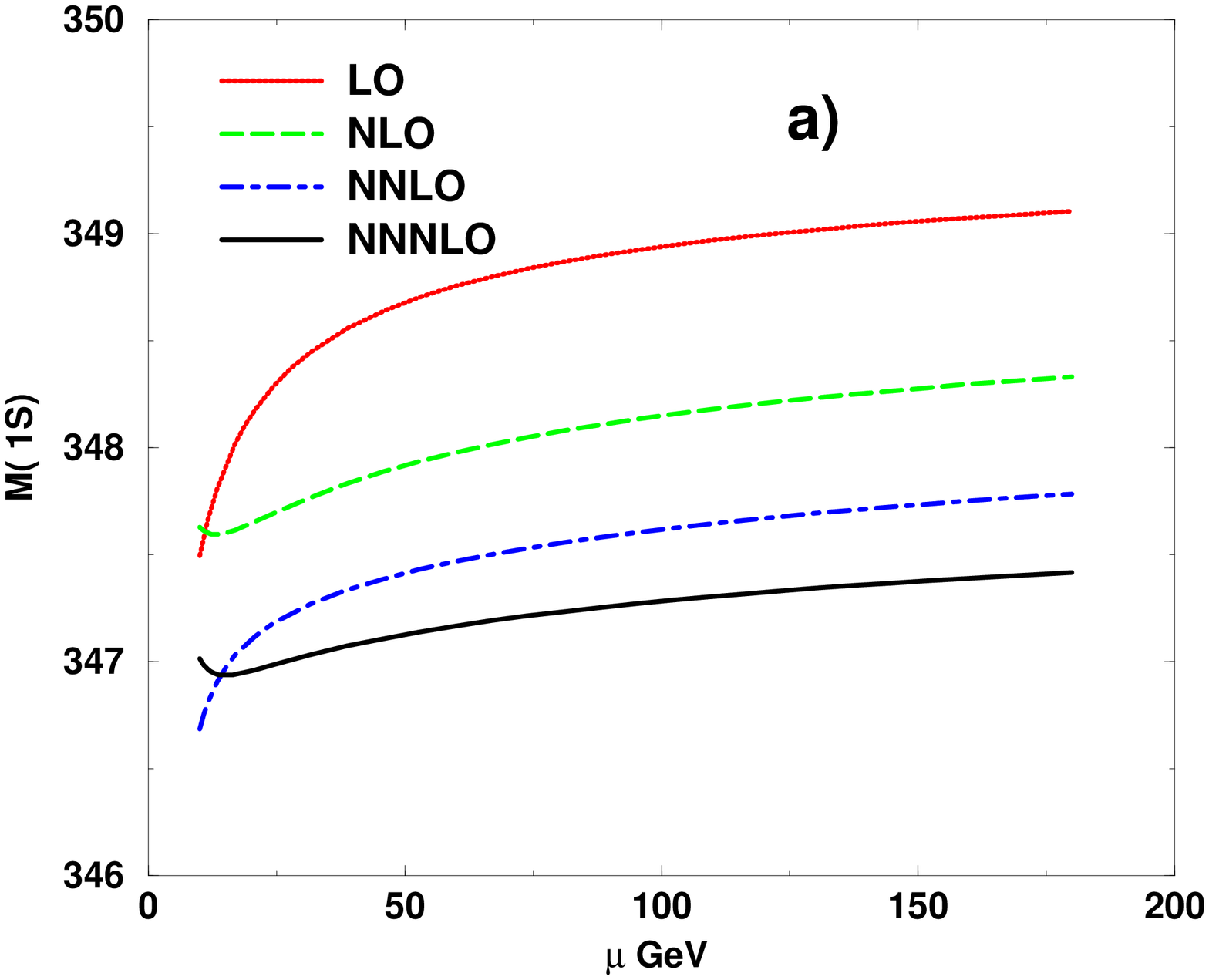} 
&
\includegraphics[width=6cm,angle=-90]{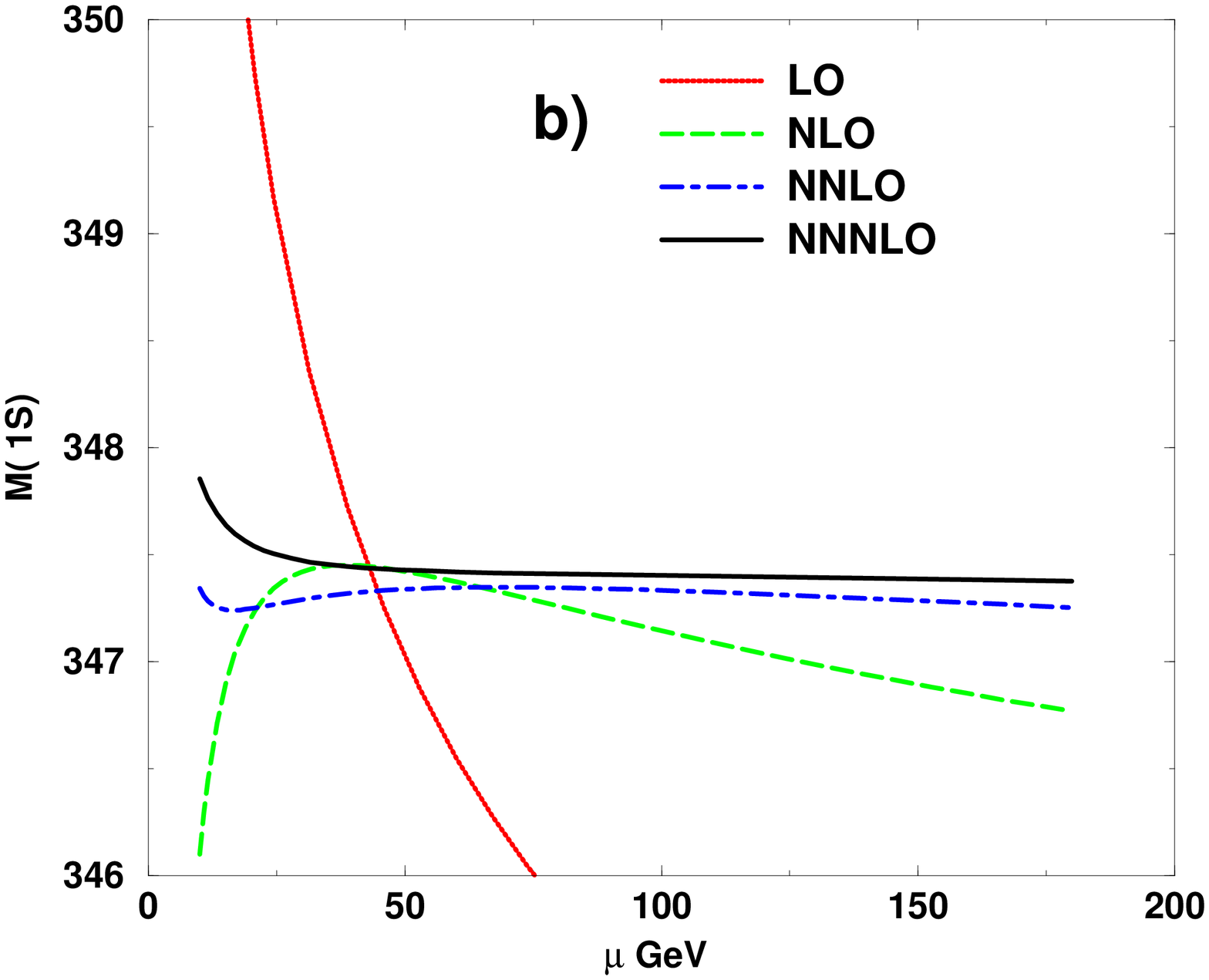}
\end{tabular}
\end{center}
\vspace*{-.5cm}
\caption{\footnotesize 
The renormalization scale dependences of the energy level of the
vector toponium $1S$ state 
for a) the pole mass scheme  and b) $\overline{\rm MS}$ 
mass scheme, respectively. 
The solid curves are the NNNLO results,
dotted, dashed and dot-dashed curves 
denote the LO, NLO and NNLO results, respectively. 
}
\label{scale-dep}
\end{figure}

\begin{figure}
\begin{center}
\includegraphics[width=8cm,angle=-90]{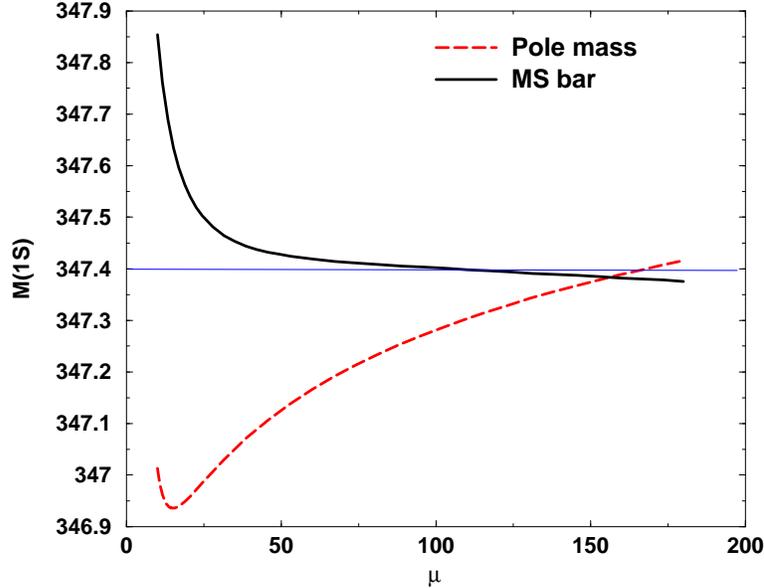}
\end{center}
\vspace*{-.5cm}
\caption{\footnotesize 
The $1S$ energy levels of the vector toponium state at NNNLO are plotted
in the pole and $\overline{\rm MS}$ schemes. 
A horizontal line, $M_{1S}=347.4$ GeV, is drawn for
a guide.
}
\label{pole-vs-MS}
\end{figure}

In Figs.~\ref{scale-dep}a) and b),
we show the renormalization scale ($\mu$) 
dependences of the $1S$ energy level when we use the pole mass and the
$\overline{\rm MS}$ mass\footnote{
The pole-$\overline{\rm MS}$ mass relation is known up to 3 loops presently.
The 4-loop correction is replaced by its large-$\beta_0$ approximation
in our analysis.
},
respectively, to express the energy level.
We used the $\epsilon$-expansion \cite{hlm} to cancel renormalons in the
$\overline{\rm MS}$ mass scheme; the relevant formulas are 
given in the Appendix.
Fig.~\ref{scale-dep}a) is essentially a reproduction of Fig.~2(b) of \cite{Penin-Steinhauser}, 
by including the leading order (LO) curve in addition.
As pointed out by \cite{Penin-Steinhauser},
the next-to-next-to-next-to-leading order
(NNNLO) prediction becomes insensitive to the scale variation at
$\mu \simeq 15$~GeV, and that the sum of the \a5 and \la5 corrections
becomes small around this scale.
On the other hand,
in Fig.~\ref{scale-dep}b), we see a good convergence behavior at 
$\mu \sim 50$--80~GeV.
In Fig.~\ref{pole-vs-MS} the vertical scale is magnified and the scale
dependences of the energy levels at the NNNLO in both mass schemes
are compared.
We find a much better stability of the prediction in the 
$\overline{\rm MS}$ mass
scheme over a wide region $40~{\rm GeV} < \mu < 160~{\rm GeV}$.
From this analysis, we consider the scale $\mu \sim 50$--80~GeV
to be an optimal scale choice in the $\overline{\rm MS}$ mass scheme.
By varying $\mu$ between 30--160~GeV, 
we estimate the theoretical error of the $\overline{\rm MS}$ mass to be order
40~MeV at NNNLO if it is extracted from the $1S$ energy level.
(We obtain an error of about 200~MeV if a similar estimate is applied for
the pole mass.)

In the pole mass scheme, it is natural to choose the renormalization scale
around the Bohr scale,
$\mu \sim C_F \alpha_S m \sim 30$~GeV.
This is because there is only one logarithm
$\log[\mu/(C_F\alpha_S m)]$ in the energy level,
associated with the renormalization scale $\mu$,\footnote{
At NNNLO, the logarithm associated with the ultrasoft scale 
$\sim \alpha_S^2 m$ is not accompanied by the renormalization scale $\mu$.
}
and because this logarithm is minimized around the Bohr scale.
On the other hand, in the $\overline{\rm MS}$ mass scheme,
two types of logarithms 
$\log(\mu/\overline{m})$ and $\log[\mu/(C_F\alpha_S \overline{m})]$ 
are included in the expression for the energy level,\footnote{
This stems from the fact that one needs to expand the
pole mass and the binding energy in the same coupling constant
$\alpha_S(\mu)$
in order to achieve the decoupling of infrared degrees of freedom
at each order of the perturbative expansion.
} 
where
$\overline{m}=m_{\overline{\rm MS}}(m_{\overline{\rm MS}})$ is the 
renormalization-group invariant $\overline{\rm MS}$ mass;
see Eqs.~(\ref{m-pole-mu}), (\ref{E_MS}) and (\ref{M1S}).
Therefore, a natural scale, which minimizes the logarithmic contributions, 
lies between the Bohr scale and
the hard scale,
$C_F \, \alpha_S  \, m < \mu < m$.
This aspect of the renormalization scale, when the leading
renormalon uncertainty is removed,  
has been discussed already for the bottomonium energy levels 
\cite{Brambilla-Sumino-Vairo}, and a further detailed study of 
the scale choice (in the context of the QCD potential) 
has been given in \cite{Sumino-pot}.

If we replace $c_3$ by $c_3(\mbox{large-$\beta_0$})$,
the corresponding figures to Figs.~\ref{scale-dep}a,b) 
and \ref{pole-vs-MS} look very similar;
these were shown in \cite{proceedings}.
The main observations in the analysis in the large-$\beta_0$ approximation 
were as follows \cite{Kiyo-Sumino, proceedings}:
(1) In the pole mass scheme, with any choice of the scale $\mu$,
the perturbative series of the $1S$ energy level does not show
a healthy convergence behavior, hence the level cannot be predicted
with an accuracy better than ${\cal O}(\Lambda_{\rm QCD})$.
(2) In the $\overline{\rm MS}$ mass scheme, one observes a good 
convergence of the
perturbative series in the range ${m}\, \alpha_S < \mu < {m}$, 
as well as stability of the prediction in this range.
Both of these observations still hold at the best of our present knowledge.
It is intriguing whether these features will remain valid even
when $a_3$ and the 4-loop relation between
the pole and $\overline{\rm MS}$ masses are computed fully in the future.

Let us address how the theoretical error of about 80~MeV 
for the top quark pole mass was obtained in Ref.~\cite{Penin-Steinhauser}.
It is dominated by the uncertainty induced by the error of the input
$\alpha_S(M_Z)$.
The uncertainty due to the scale dependence was estimated by varying
$\mu$ between 10--30~GeV and an error of 20.5(=41/2)~MeV was assigned as
an uncertainty from this source.
Uncertainties from other sources were estimated to be even smaller.
Here, let us concentrate on the error estimate from the scale dependence
and discuss its problem.
The smallness of this error ensures, partly, the smallness of the
total error (80~MeV).
However, if the same estimation method is
applied to the LO and NLO curves in Fig.~\ref{scale-dep}a), we should
infer that the NNLO correction is small, in contradiction
to its true large size.
Thus, apparently there is a danger in relying on this estimation method.
By contrast, our error estimate of the top quark
$\overline{\rm MS}$ mass from the scale dependence
does not suffer from the same problem.
The same estimation method works at lower orders,
because the perturbative series in Fig.~\ref{scale-dep}b)
shows a healthy convergence behavior and the
scale dependence decreases as we include more terms around the
relevant scales.

%

At this stage, there seems to be a puzzling point:
on the one hand, the validity
of the large-$\beta_0$ approximation is known to lead to an 
${\cal O}(\Lambda_{\rm QCD})$ uncertainty of the pole mass;
on the other hand, the small
size of the \a5 plus \la5 corrections in the range $\mu \sim 10$--30~GeV
appears to be incompatible with the renormalon picture.

\begin{figure}[tbp]
  \hspace*{\fill}
    \includegraphics[width=6.5cm]{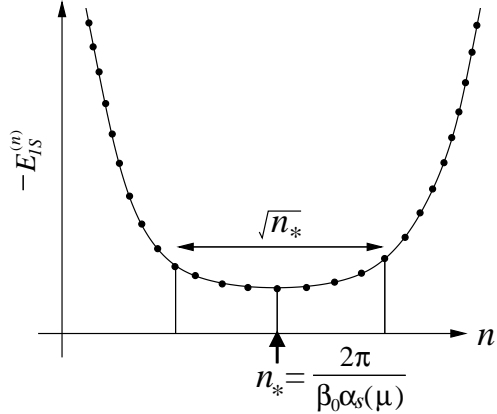}
  \hspace*{\fill}
  \\
  \hspace*{\fill}
\caption{\footnotesize
The graph showing schematically the asymptotic behavior of the $n$-th
term of $- E_{1S}$ in the large-$\beta_0$ approximation for $n \gg 1$.
      \label{asympt}
}
  \hspace*{\fill}
\end{figure}

Let us recall the estimate of the renormalon
uncertainty in the large-$\beta_0$ 
approximation (see e.g.\ \cite{Sumino-pro}).
Asymptotically the perturbative series of the $1S$ energy level, if
expressed in the pole mass, behaves as
\bea
E_{1S}^{(n)} \sim \mbox{const.} 
\times {\mu} \, \alpha_S(\mu) \times
\biggl\{ \frac{\beta_0 \alpha_S(\mu)}{2\pi} \biggr\}^n \,
\times n! 
~~~~~~
\mbox{for~~~$n \gg 1$.}
\eea
It becomes minimal at $n \approx n_* \equiv 2\pi/(\beta_0 \alpha_S(\mu))$.
The size of the term scarcely changes within the range 
$n \in ( n_* - \sqrt{n_*}, n_* + \sqrt{n_*} )$;
see Fig.~\ref{asympt}.\footnote{
Using the Stirling formula, one may easily find 
an approximate position $n_*$ of the minimum of the series.
Then, by expanding around the minimum, one finds an 
approximate form
$ n! \, n_*^{-n} \approx \sqrt{2\pi n_*} \, \exp [-n_* + (n-n_*)^2/(2n_*)]
\sim \sqrt{2\pi n_*}\, \exp (-n_* )$ in the range 
$|n - n_*| \simlt \sqrt{n_*}$.
}
We may consider the uncertainty of this asymptotic series
as the sum of the terms within this range, since we are not sure
where to truncate the series within this range:
\bea
\delta E_{1S} \sim 
\left| \sum_{ n = n_* - \sqrt{n_*} }^{n_* + \sqrt{n_*}} \, 
E_{1S}^{(n)} \right|
\sim \Lambda_{\rm QCD} .
\eea
The $\mu$-dependence vanishes in this sum, and this leads to the
claimed uncertainty.
This argument shows that when the relevant coupling constant
$\alpha_S(\mu)$ is small (corresponding scale $\mu$ is large),
$n_*$ is large.
Then each term of the series for 
$n \in ( n_* - \sqrt{n_*}, n_* + \sqrt{n_*} )$
can become considerably smaller than $\Lambda_{\rm QCD}$.

According to this argument, the small size of the \a5 plus \la5 correction
at certain scales does not generally lead to an uncertainty considerably
smaller than $\Lambda_{\rm QCD}$.
While an error estimate should necessarily be more or less subjective,
as long as the large-$\beta_0$ approximation is valid, we should at least bear in mind
how the theoretical uncertainty is estimated in this framework.
Incidentally, based on the large-$\beta_0$ approximation, the $\overline{\rm MS}$ mass extracted from the
$1S$ energy level has an uncertainty of order
$\Lambda_{\rm QCD}^3/(\alpha_S m_t)^2 \sim 
\Lambda_{\rm QCD}^3/\mu_{\rm opt}^2 \sim 3$--10~MeV
originating from the next-to-leading order renormalon contribution
\cite{al}.
Thus, the above perturbative error of order 40~MeV is still significantly
larger than this contribution.

To conclude, we observe a much more stable prediction of the
toponium $1S$ energy level when we use the $\overline{\rm MS}$ mass instead of the pole
mass.
Considering this situation and the good agreement of the 
large-$\beta_0$ approximation with the presently known
corrections, we consider a theoretical uncertainty of the pole mass
of order $\Lambda_{\rm QCD}$$\sim 200$--300~MeV to be legitimate.
On the other hand, based on the argument in \cite{Kiyo-Sumino}, it is likely
that the top quark $\overline{\rm MS}$ mass can be extracted with an accuracy 
of order 40~MeV, 
once the 4-loop relation between the pole and $\overline{\rm MS}$ mass is
calculated.
This number may be compared with the most recent estimate 
\cite{Martinez-Miquel-2002} of the
experimental error (including some systematic errors) of 
19~MeV in the determination of the top quark $1S$ mass, 
corresponding to a 3-parameter fit with 
an integrated luminosity of 300~fb$^{-1}$.

\section*{Acknowledgement}
This work was completed through discussion during the Quarkonium
Working Group Workshop held at CERN in Nov.\ 2002.
We are grateful to those who participated in the discussion 
and in particular to the organizers of this workshop.
Y. K. was supported by the Japan Society for the Promotion of Science.

\section*{Appendix}

In this appendix we list the formulas we use to convert the
energy level of the quarkonium $1S$ state from the pole mass scheme to the 
$\overline{\rm MS}$ mass scheme using the $\varepsilon$-expansion \cite{hlm}. 

The energy of the quarkonium $1S$ state is given by 
\begin{eqnarray}
&&
M_{1S}
=
2 m_{\rm pole} + E_{1S}(m_{\rm pole},\alpha_S(\mu))
\end{eqnarray}
as a function of $m_{\rm pole}$ and $\alpha_{S}(\mu)=\alpha_S^{(n_f)}(\mu)$ 
in the pole mass scheme, where $n_f$ is the number of light quark flavors
($n_f=4,\, 5$ for the bottomonium and toponium, respectively).
Mass relation between the pole and $\overline{\rm MS}$ masses is given by 
\begin{eqnarray}
&& \hspace{-1cm}
m_{\rm pole}
=
\overline{m}\,
\left\{ 1 + d_0 \frac{\varepsilon \alpha_S(\overline{m})}{\pi}
         +  d_1 \left(\frac{\varepsilon \alpha_S(\overline{m})}{\pi}\right)^2
         +  d_2  \left(\frac{\varepsilon \alpha_S(\overline{m})}{\pi}\right)^3
         +  d_3  \left(\frac{\varepsilon \alpha_S(\overline{m})}{\pi}\right)^4
         + {\cal O} (\varepsilon^5 )
\right\}\, ,
\label{m-pole}
\end{eqnarray}
where $\varepsilon=1$ is the expansion parameter in the $\varepsilon$-expansion, 
$\overline{m} \equiv m_{\overline{\rm MS}}(m_{\overline{\rm MS}})$,\, $d_0=4/3$.
The coefficients $d_1$ and $d_2$ are obtained from the 2-loop \cite{2loop-mass-relation}
and 3-loop \cite{3loop-mass-relation} mass relations 
\footnote{
The same relation was obtained numerically before
in \cite{Chetyrkin-Steinhauser} in a certain approximation.
}, 
respectively, by rewriting them in terms of the coupling of the theory with
$n_f$ flavors only. These are given by
\bea
d_1 &=&
{\frac{307}{32}} + {\frac{{{\pi }^2}}{3}} 
      + {\frac{{{\pi }^2}\,\log 2}{9}} - {\frac{\zeta_3}{6}}
+ 
  n_f\,\left( -{\frac{71}{144}} - {\frac{{{\pi }^2}}{18}} \right)
\nonumber \\ &\simeq&
13.4434-1.04137 \, n_f ,
\\ ~~~ \nonumber \\
d_2 &=&
{\frac{8462917}{93312}} + {\frac{652841\,{{\pi }^2}}{38880}} - 
  {\frac{695\,{{\pi }^4}}{7776}} - {\frac{575\,{{\pi }^2}\,\log 2}{162}} 
\nonumber \\ &&
- 
  {\frac{22\,{{\pi }^2}\,{{\log^2 2}}}{81}} - 
  {\frac{55\,{{\log^4 2}}}{162}} 
- 
  {\frac{220\,{\rm Li}_4(\frac{1}{2})}{27}} 
+ {\frac{58\,\zeta_3}{27}} - 
  {\frac{1439\,{{\pi }^2}\,\zeta_3}{432}} + 
  {\frac{1975\,\zeta_5}{216}}
\nonumber \\ &&
+ 
  n_f\,\left( -{\frac{231847}{23328}} - 
     {\frac{991\,{{\pi }^2}}{648}} + {\frac{61\,{{\pi }^4}}{1944}} - 
     {\frac{11\,{{\pi }^2}\,\log 2}{81}} + 
     {\frac{2\,{{\pi }^2}\,{{\log^2 2}}}{81}} + 
     {\frac{{{\log^4 2}}}{81}} 
\right.
\nonumber \\ &&
\left.
+ 
     {\frac{8\,{\rm Li}_4(\frac{1}{2})}{27}} - 
     {\frac{241\,\zeta_3}{72}} \right)  + 
  {n_f^2}\,\left( {\frac{2353}{23328}} + 
     {\frac{13\,{{\pi }^2}}{324}} + {\frac{7\,\zeta_3}{54}}
     \right)  
\nonumber \\ &\simeq &
190.391 - 26.6551\, n_f +  0.652691\, n_f^2  \, ,
\eea
with $\zeta_3=1.20206\cdots,\, {\rm Li}_4(\frac{1}{2})=0.517479\cdots $. 
The third coefficient $d_3$ is not known exactly yet.
In this paper we use its value in the large-$\beta_0$ approximation 
\cite{bb}:
\bea
d_3(\mbox{large-}\beta_0) &=&
\frac{ \beta_0^3}{64}\,
\left( {\frac{42979}{5184}} + {\frac{89\,{{\pi }^2}}{18}} + 
       {\frac{71\,{{\pi }^4}}{120}} + 
       {\frac{317\, \zeta_3}{12}}\
\right)
\nonumber \\ &\simeq & 
3046.29 - 553.872\,{n_f} +   33.568\,{{{n_f}}^2} - 0.678141\,{{{n_f}}^3}.
\eea
To achieve the renormalon cancellation between $2\, m_{\rm pole}$ and 
$E_{1S}(m_{\rm pole},\alpha_S(\mu))$ order by order in the $\varepsilon$-expansion, 
we must use the same coupling constant $\alpha_S(\mu)$ in the series
expansions of 
$2m_{\rm pole}$ and $E_{1S}$. 
Therefore, $\alpha_S(\overline{m})$ is re-expressed in 
terms of $\alpha_S(\mu)$ as 
\begin{eqnarray}
&&
\alpha_S(\overline{m})
=
\alpha_S(\mu)
\left\{ 1 
+ \frac{\beta_0 \, \log \left(\frac{\mu}{\overline{m}}\right) }{2} \, 
  \left( \frac{\varepsilon \, \alpha_S(\mu)}{\pi} \right)
+ \left( \frac{\beta_1 \, \log\left(\frac{\mu}{\overline{m}}\right)}{8}
        +\frac{\beta_0^2 \, \log^2 \left(\frac{\mu}{\overline{m}}\right)}{4} \right)
  \left(  \frac{\varepsilon \,\alpha_S(\mu)}{\pi} \right)^2
\right.
\nonumber \\
&& ~~~~~~~~~~
\left.
+ \left( \frac{\beta_2 \, \log \left(\frac{\mu}{\overline{m}}\right)}{32}
       + \frac{5\beta_0 \, \beta_1 \, \log^2 \left(\frac{\mu}{\overline{m}}\right)}{32}
       +\frac{\beta_0^3 \, \log^3 \left(\frac{\mu}{\overline{m}}\right)}{8}  
 \right)   \left(  \frac{\varepsilon \,\alpha_S(\mu)}{\pi} \right)^3
+{\cal O}(\varepsilon^4)
\right\} \, ,
\label{a-m}
\end{eqnarray}
using the coefficients of the QCD $\beta$-function:
\begin{eqnarray}
&& 
\beta_0 = 11 - {\frac{2\,{n_f}}{3}} \, , ~~~~~    
\beta_1 = 102 - {\frac{38\,{n_f}}{3}} \, , ~~~~~        
\beta_2 =  {\frac{2857}{2}} - {\frac{5033\,{n_f}}{18}} + {\frac{325\,{{{n_f}}^2}}{54}}. 
\end{eqnarray}
Using Eqs. (\ref{m-pole}) and (\ref{a-m}), we obtain the $\varepsilon$-expansion 
for $m_{\rm pole}$ in terms of $\alpha_S(\mu)$, 
\bea
m_{\rm pole}
&=&
\overline{m} 
\times
\left( 1 + \sum_{n=1}^{4}\, \widetilde{d}_{n-1}(l_\mu) \, 
            \varepsilon^{n} \left(\frac{\alpha_S(\mu)}{\pi}\right)^{n} 
\right)
+{\cal O}(\varepsilon^5) \, ,
\label{m-pole-mu}
\end{eqnarray}
where the coefficients $\widetilde{d}_{n}(l_\mu)$ are 
functions of $l_\mu=\log(\mu/\overline{m})$ which enter via Eq. (\ref{a-m}).

The binding energy $E_{1S}(m_{\rm pole}, \alpha_S(\mu))$ is given by
\bea
&&
E_{1S}
=
-
\frac{4}{9} \alpha_S(\mu)^2 \, m_{\rm pole} \sum_{n=0}^\infty 
\varepsilon^{n+1}
\left( \frac{\alpha_S(\mu)}{\pi} \right)^n P_n(L_{\mu})\, , 
\eea
where $L_{\mu}=\log\left[ \mu/(C_F \alpha_S(\mu) m_{\rm pole}) \right]$, 
and $P_n(L_{\mu})$  are given by 
\bea
&& P_0(L_\mu) = 1\, , ~~~~~~P_1(L_\mu) = \beta_0  \, L_\mu + c_1 ,
\nonumber \\
&& P_2 (L_\mu) = 
\frac{3}{4} \beta_0^2 \, {{L_\mu}^2} + 
  \left( - \frac{1}{2} \beta_0^2 
             + 
     {\frac{1}{4}\beta_1 } + 
     {\frac{3}{2} \beta_0 {c_1}} \right) {L_\mu}  + 
  {c_2} ,
\nonumber \\
&& P_3(L_\mu) =
\frac{1}{2}\beta_0^3  \, {L_\mu}^3 + 
  \left( -\frac{7}{8}\beta_0^3
        + \frac{7}{16}\beta_0\beta_1
          + 
     \frac{3}{2}\beta_0^2c_1 \right) {L_\mu}^2
\nonumber \\ 
&& ~~~
      + \left( \frac{1}{4}\beta_0^3
        - \frac{1}{4}\beta_0\beta_1
          + 
     \frac{1}{16}\beta_2 - 
     \frac{3}{4}\beta_0^2c_1 + 
     \frac{3}{8}\beta_1c_1 + 
     2\beta_0c_2 \right){L_\mu}  + c_3 \, ,
\eea
with
\bea
&& c_1 = \frac{97}{6}-\frac{11}{9}{n_f} ,
\nonumber \\
&& c_2 = 
\frac{1793}{12} + \frac{2917\pi^2}{216}
-\frac{9\pi^4}{32}+\frac{275 \zeta_3}{4} 
+ \Bigl( - \frac{1693}{72}-\frac{11\pi^2}{18}-\frac{19\zeta_3}{2} 
\Bigr){n_f}
+ \Bigl( \frac{77}{108}+\frac{\pi^2}{54}+\frac{2\zeta_3}{9} \Bigr)
{n_f}^2 .
\nonumber \\
\eea
The terms which contain $L_{\mu}$ are determined by renormalization-group equation
from lower order constants $c_{1,\,2}$. The $c_{1,\,2}$ are taken from 
\cite{Pineda-Yndurain}, $c_3$ is given in Eq.~(\ref{full}). 
To obtain the $\varepsilon$-expansion in the $\overline{\rm MS}$ scheme,
we re-express the pole mass in $E_{1S}(m_{\rm pole}, \alpha_S(\mu))$ by 
$\overline{m}$ and $\alpha_S(\mu)$ employing the mass relation Eq.~(\ref{m-pole-mu}),
which gives 
\bea
E_{1S}=-\frac{4}{9}  \alpha_S(\mu)^2 \, \overline{m}
\sum_{n=0}^3 \varepsilon^{n+1}
\left( \frac{\alpha_S(\mu)}{\pi} \right)^n \widetilde{P}_n(\widetilde{L}_{\mu}, l_\mu)\, , 
+{\cal O}(\varepsilon^5) \, ,
\label{E_MS}
\eea
with $\widetilde{L}_{\mu}= \log \left[ \mu/ (C_F \alpha_S (\mu) \overline{m}  ) \right] $. 
Using the $\varepsilon$-expansions Eqs.~(\ref{m-pole-mu}) and  (\ref{E_MS}), 
$M_{1S}$ is rewritten as 
\bea
M_{1S} 
&=& 
2\overline{m}
\left(1+\sum_{n=1}^{4}\, 
        \widetilde{d}_{n-1}(l_\mu) \, 
        \varepsilon^{n} 
       \left(\frac{\alpha_S(\mu)}{\pi}\right)^{n}
\right)
-
        \frac{4}{9}  
        \alpha_S(\mu)^2 \, \overline{m}
        \left(
        \sum_{n=0}^3 
        \varepsilon^{n+1}
        \left( \frac{\alpha_S(\mu)}{\pi} \right)^n 
        \widetilde{P}_n ( \widetilde{L}_{\mu}, l_\mu)
        \right)
\nonumber \\
&=&
       2 \overline{m}
\left[1+\sum_{n=1}^{4}\, 
       \left( \widetilde{d}_{n-1}(l_\mu) 
            - \frac{2 \pi \alpha_S(\mu)}{9}  \,
               \widetilde{P}_{n-1} ( \widetilde{L}_{\mu}, l_\mu)
       \right)
       \times
      \left(\frac{ \varepsilon\, \alpha_S(\mu)}{\pi}\right)^{n}
\right]\, .
\label{M1S}
\eea
Setting the expansion parameter $\varepsilon=1$ in the final expression, 
the $n$-th order
correction to $M_{1S}$ in the $\overline{\rm MS}$ scheme 
is given by
$
2 \overline{m} \times 
\left(\alpha_S(\mu)/\pi\right)^{n}
\, 
\left[ 
  \widetilde{d}_{n-1} (l_\mu)
- \left(2 \pi \alpha_S(\mu)/9\right) \, 
  \widetilde{P}_{n-1} ( \widetilde{L}_{\mu}, l_\mu)
\right] 
$.


\end{document}